\documentclass[manuscript]{acmart}

\AtBeginDocument{%
  }

\usepackage{tabu}                      
\usepackage{booktabs}                  
\usepackage{lipsum}                    
\usepackage{mwe}                       
\usepackage{multirow}
\usepackage{cleveref}
\usepackage{mathptmx}                  
\usepackage{tabularx}
\usepackage{multirow}
\usepackage{tabularx}
\usepackage{cleveref}
\usepackage{makecell}

\setcopyright{acmlicensed}
\copyrightyear{2018}
\acmYear{2018}
\acmDOI{XXXXXXX.XXXXXXX}
\acmConference[ISS '26]{Make sure to enter the correct
  conference title from your rights confirmation email}{June 03--05,
  2018}{Woodstock, NY}
\acmISBN{978-1-4503-XXXX-X/2018/06}

\begin{document}

\title{A Taxonomy of Confabulations and the Perception-Reality Gap in LLM-Assisted Immersive Scene Editing}

\author{Junlong Chen}
\correspondingauthor
\affiliation{%
  \institution{University of Cambridge}
  \city{Cambridge}
  \country{United Kingdom}
}
\email{jc2375@cam.ac.uk}
\orcid{0000-0002-7375-6525}

\author{Per Ola Kristensson}
\affiliation{%
  \institution{University of Cambridge}
  \city{Cambridge}
  \country{United Kingdom}}
\email{pok21@cam.ac.uk}
\orcid{0000-0002-7139-871X}

\renewcommand{\shortauthors}{Chen et al.}

\begin{abstract}
  Large language models (LLMs) are being increasingly integrated into immersive environments and design workflows, providing application prospects in areas such as rapid scene prototyping for non-expert users and scene understanding capabilities for accessibility design. While many workflows that incorporate LLMs in immersive spaces are proposed, such systems can exhibit errors, potentially resulting in frustration, loss of user trust, and compromised user safety. This paper studies the underexplored area of LLM confabulations in immersive 3D scene editing contexts. Through an exploratory study with 24 non-expert users, we construct a taxonomy of the different types of confabulation observed in LLM-assisted immersive 3D scene editing. We report their prevalence and disruptiveness, and define the construct \emph{perception-reality gap} to help understand the gap between the actual and perceived occurrence of confabulations. We highlight the observed saturation of confabulation awareness under load and conclude by discussing design implications for confabulation mitigation in future LLM-assisted systems in immersive 3D scenes.
\end{abstract}

\ccsdesc[500]{Human-centered computing~Empirical studies in HCI}
\ccsdesc[500]{Human-centered computing~Empirical studies in interaction design}
\ccsdesc[500]{Human-centered computing~Virtual reality}


\begin{teaserfigure}
  \centering
  \includegraphics[width=0.33\textwidth]{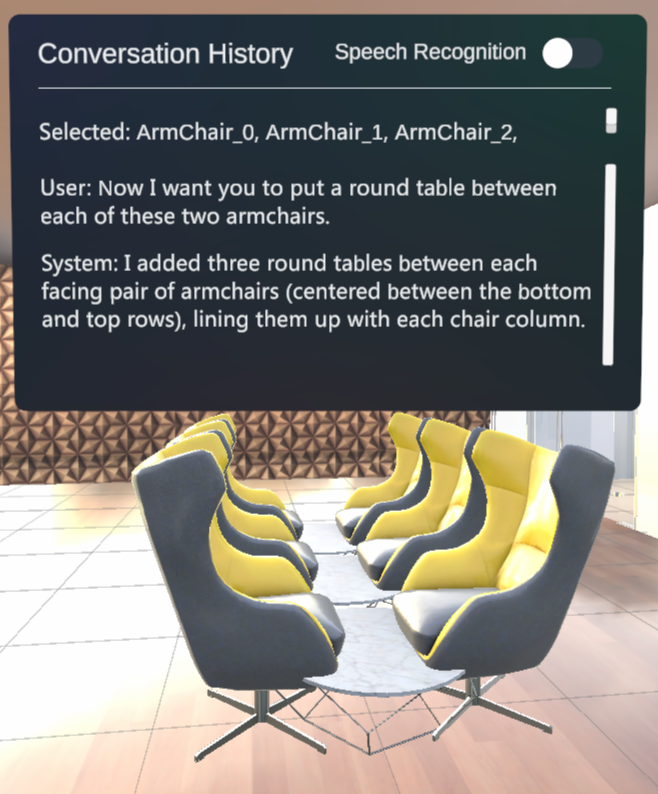}
  \includegraphics[width=0.33\linewidth]{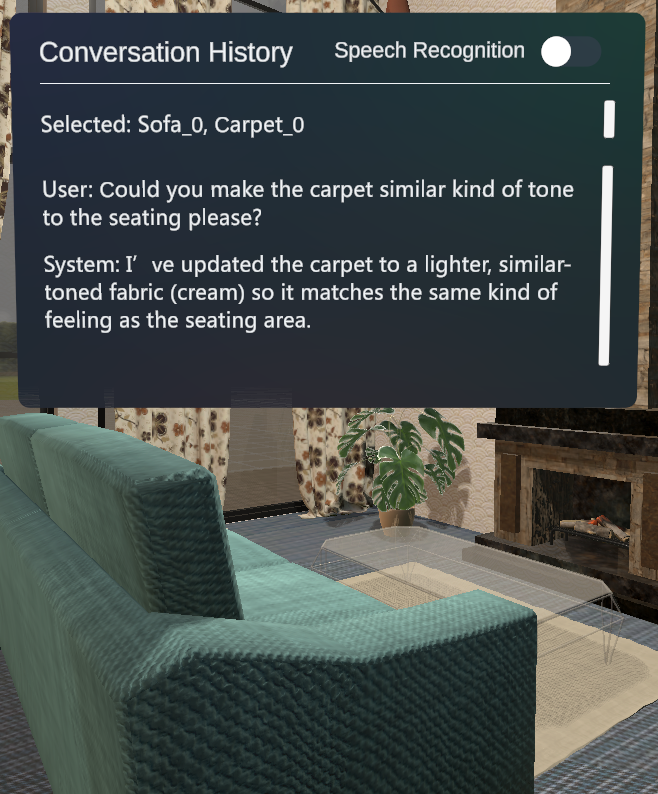}
  \includegraphics[width=0.33\linewidth]{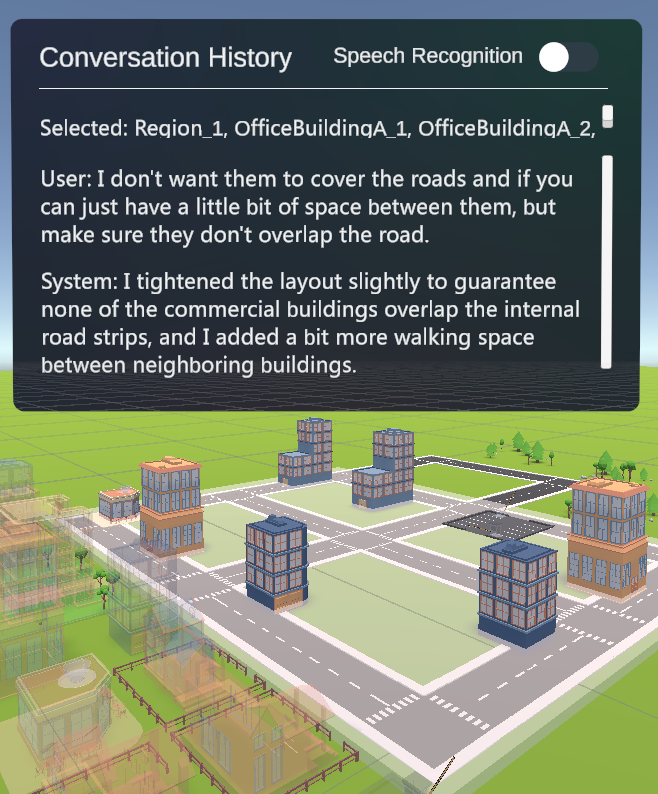}
  \caption{Examples of different types of confabulations identified through an empirical user study of LLM-assisted immersive scene editing workflows. Left: Spatial/placement error with armchairs overlapping with round tables in the office scene. Middle: Interpretation error with incorrect carpet appearance edit in the loft scene. Right: Spatial/placement error with buildings overlapping with roads in the city scene.}
  \label{fig:teaser}
\end{teaserfigure}


\maketitle

\section{Introduction}

Generative Artificial Intelligence (GenAI) models have widespread applications in processing text and images~\cite{wu2023multimodal, huang2023language}, and are increasingly showing their potential in generating 3D scene information. These workflows typically include pipelines that alter pixels or neural representations such as vision-language models (VLMs) for scene generation and editing~\cite{sun2025layoutvlm, huang2024blenderalchemy}, as well as 3D Gaussian Splatting (3DGS) frameworks for scene editing based on textual and/or image input~\cite{chen2024gaussianeditor, shu2025gaussedit}. Alternatively, there are also symbolic pipelines that edit structural data which uniquely define a 3D scene~\cite{zheng2025editroom, chen2025llmer, yang2025llm}. While generative models provide high scene editing quality, they lack deterministic scene representations which makes it difficult to integrate them with downstream pipelines for different applications such as accessibility support, CAD tools, and physics simulation with high precision requirements. As a result, JSON-in JSON-out scene editing pipelines~\cite{chen2025llmer, yang2025llm} have emerged to perform structured edits to 3D scenes. Throughout this paper, we define LLM-assisted immersive 3D scene editing as such JSON-in, JSON-out editing workflows, where 3D scene information is often encoded in the form of scene graphs and are integrated into virtual and augmented reality (VR/AR) to facilitate 3D scene editing~\cite{tang2025llm, de2024llmr, chen2025llmer, chen2025analyzing, zhang2024vrcopilot}.

However, similar to LLMs and multimodal large language models (MLLMs), LLM-assisted 3D scene editing workflows can also exhibit hallucinations and produce confabulated content~\cite{chen2025analyzing}. Recent works have explored how LLMs~\cite{ji2023survey, huang2025survey, zhang2025siren} and MLLMs~\cite{sahoo2024comprehensive, liu2024survey, bai2024hallucination} can produce confabulated content, and many works have proposed mitigation strategies~\cite{zhang2025llm, tonmoy2024comprehensive, bai2024hallucination, huang2025survey}. While many works have explored various opportunities of integrating LLMs in 3D scene editing workflows, a research gap exists in identifying confabulations\footnote{We clarify that the term \emph{confabulation} is used in this paper to describe cases where the LLM-assisted scene editing system produces outcomes which conflict with the user's instruction, or issues edits which are inconsistent with common sense of 3D scenes, such as generating overlapping objects or other physically-impossible object positions. In contrast to the majority of the existing literature, which uses the term ``hallucination'' to describe erroneous output of LLMs, we use the term \emph{confabulation} as this term does not imply the LLM system has consciousness and thus provides a more accurate description of the underpinning mechanism generating plausible-sounding but incorrect edits without awareness of the error~\cite{smith2023hallucination}.} and failure cases, which motivates our work.

This paper aims to address the following research questions (RQs):
\begin{itemize}
    \item \textbf{RQ1:} What representative types of confabulations exist in LLM-assisted immersive scene editing workflows and how does their prevalence vary across different task types?
    \item \textbf{RQ2:} How disruptive are confabulations in different tasks, and how do they affect user perceived trust and workload?
    \item \textbf{RQ3:} How well do users perceive different confabulation types relative to their actual occurrence, and what factors influence this \emph{perception-reality gap}\footnote{We introduce the \emph{perception-reality gap} in AI for extended reality (XR) and define it as the discrepancy between the number and types of confabulations users believe they have encountered versus those that actually occurred to help understand how user awareness of LLM failures breaks down under realistic task conditions and inform the design of system-side mitigation strategies.}?
    \item \textbf{RQ4:} What are the design implications for future confabulation mitigation strategies within the context of LLM-assisted scene editing workflows and beyond?
\end{itemize}

To study these research questions, we build an LLM-assisted 3D scene editing workflow and apply it to different scenes. Through an empirical user study, we invite 24 participants to complete three types of scene editing tasks, ranging from appearance edits to edits of object position and orientation, and high-level edits of large regions based on functional requirements. 

Based on interaction data and post-experience questionnaire results from the user study, we construct a taxonomy of nine confabulation types based on inductive open coding of the actual log data. 
We report participant ratings to discuss whether certain confabulations were more disruptive than others, and how confabulations affected user perceived trust and overall load across different task types.
We also compare actual confabulations against confabulations reported by users (perceived confabulations) to discuss whether certain types of confabulations were more difficult to notice and characterize the perception-reality gap.

Based on this empirical evidence, we highlight several counter-intuitive dynamics unique to interactive immersive environments. First, confabulation awareness saturates under higher cognitive load. In high-error environments (such as the office and city scenes), users' ability to track anomalies breaks down, causing them to systematically underestimate error rates. Consequently, human oversight cannot be relied upon as a sole detection mechanism. Second, the immersive medium creates a perceptual asymmetry between spatial and semantic failures. While explicit spatial violations (e.g., collisions and out-of-boundary placements) are readily identified, semantic interpretation errors frequently go unnoticed. Third, task framing affects user tolerance toward errors. For example, stylistic design tasks (the loft scene) foster significantly higher user trust and lower perceived disruption than functional design tasks governed by strict spatial coordinates (the city scene), suggesting that task-level context can act as a design lever to help increase user trust and reduce perceived disruption. Fourth, highly confident textual responses from the LLM frequently mask incorrect system executions (such as false claims of task completion or silent action failures), which calls for automated, independent verification of structural states rather than relying on textual assurances. 

Finally, we translate these empirical insights into actionable design implications. These guidelines are structured to help system designers target the most prevalent and disruptive confabulations to reduce cognitive load and enhance user trust in future co-creative spatial workflows.


\section{Related Work}

\subsection{LLM-Assisted Immersive Scene Editing}

LLMs are increasingly adopted in immersive interactive systems, and many works have studied the powerful new capabilities brought by this integration~\cite{tang2025llm, hirzle2023xr}. In a scoping review by Hirzle et al.~\cite{hirzle2023xr}, a large percentage of works focused on the integration of AI and XR to support scene creation (37.6\%) and interaction (20.3\%). 
Within the context of AI for XR scene editing, a typical workflow is to represent immersive 3D scenes using scene graphs in JSON format to allow LLMs to process 3D scene information~\cite{de2024llmr, chen2025llmer, chen2025analyzing, zhang2024vrcopilot}. Based on this concept, various LLM-assisted immersive scene editing workflows have been proposed. While most approaches represent scenes in JSON format scene graphs and feed this scene information to the AI model, the output of AI models slightly differ, which support slightly different downstream tasks.

One approach in existing literature is to instruct the LLM agent to generate an updated scene layout in a structured format to connect to post-processing scripts to update the scene. A typical example is LLplace~\cite{yang2025llplace} where a Low-Rank Adaptation (LoRA) fine-tuned LLM takes scene information encoded in JSON as input an outputs an updated scene layout in JSON format as well to generate and edit 3D scene layouts. SceneReVis~\cite{zhao2026scenerevis} also applied this `JSON-in, JSON-out' approach to iteratively resolve spatial conflicts and edit 3D scenes via an agentic reinforcement learning pipeline. LayoutGPT~\cite{feng2023layoutgpt} and LLMER~\cite{chen2025llmer} also prompt LLMs to generate style sheet language and JSON data respectively, which are parsed by downstream modules for 3D scene synthesis and interactive XR content generation.

Another approach is to instruct LLMs to generate code to be compiled to control scene elements and dynamic behavior. LLMR~\cite{de2024llmr} adopts this approach and leverages multiple generative pre-trained transformers (GPTs) to complete specialized tasks to plan, analyze the scene, manage skills, and build and inspect code to update the scene. DreamCodeVR~\cite{giunchi2024dreamcodevr}, MagicItem~\cite{kurai2025magicitem}, and SceneCraft~\cite{hu2024scenecraft} also use LLM agents to generate code to define virtual object behaviors and optimize object spatial relations to edit the scene in real time. 

While many works have explored the integration of LLMs in immersive scene editing workflows, the consequences of LLM confabulations within this application context remains little explored. Only a few works studied spatial hallucination~\cite{peng2025understanding}, but these existing works are either not based on actual user interaction data or not studied within immersive contexts. This motivates our work to address this research gap.



\subsection{Confabulations in LLMs and MLLMs}
The term ``hallucination'' has been widely used for natural language generation (NLG) tasks to refer to textual responses generated by AI that is nonsensical or unfaithful to the source input~\cite{ji2023survey}. Ji et al.~\cite{ji2023survey} categorize them into \textit{intrinsic hallucinations} which refer to cases where generated output contradicts the source content, and \textit{extrinsic hallucinations} where the generated output can neither be supported or contradicted by the source content. Huang et al.~\cite{huang2025survey} refined this taxonomy and suggested categorizing hallucinations into \textit{factuality hallucinations} (factual inconsistencies of generated content compared to verifiable real-world facts) and \textit{faithfulness hallucinations} (divergence of generated content from user input or previously generated content).

Hallucinations have also been studied for multimodal large language models (MLLMs). MLLMs~\cite{liu2023visual, zhu2023minigpt} refers to models which enable LLMs to interpret data from different modalities. Bai et al.~\cite{bai2024hallucination} classify hallucination in MLLMs into category hallucination (identifying non-existent or incorrect object categories from a given image), attribute hallucination (hallucinated descriptions of object attributes), and relation hallucination (hallucinated relationships among objects). They classify causes of these hallucinations into hallucinations from data, from the model, from the training process, and from the inference process. Liu et al.~\cite{liu2024survey} summarized different causes and mitigation methods of hallucinations in large vision-language models (LVLMs). These hallucination causes are classified based on where they occur in the LVLM pipeline, which includes hallucinations due to data (bias in data or irrelevant annotations), the vision encoder (such as limited visual resolution), the connection module (such as constraints on the tokens), and the LLM (such as insufficient attention in the context). Li et al.~\cite{li2023evaluating} and Zhou et al.~\cite{zhou2023analyzing} highlighted \textit{object hallucinations} in LVLMs, where the model generates hallucinated objects in the textual description which are not present in the given ground-truth image.


A large volume of works have discussed the severe negative consequences of LLM hallucinations in various fields. These involve, for example, producing false details within clinical contexts~\cite{omar2025multi}, accepting incorrect assumptions in legal practice~\cite{dahl2024large}, and inventing non-existent libraries in software development~\cite{twist2025library}. Xu et al.~\cite{xu2024hallucination} claim that hallucination is an innate limitation of large language models and it is impossible to eliminate hallucination in real-world LLMs. This calls for robust mechanisms to ensure the safety of LLM-based systems to mitigate the negative consequences discussed above. The risks of LLM hallucinations in 3D spaces are being increasingly studied~\cite{peng2025understanding, yang20253d, wang2024hallo3d}, but there is still a research gap in studying confabulation cases in LLM-assisted \textit{immersive} scene editing workflows to inform future design of guardrails, which motivates this work.




\subsection{Human-AI Collaboration}

As AI-assisted tools become more prevalent in workflows, an increasing number of works have highlighted patterns and factors to consider when designing human-AI collaborative systems due to issues in user trust, user agency, and model transparency.

In terms of user trust, recent studies reveal both a lack of user trust in AI systems~\cite{bach2024systematic} and an emerging tendency of users to over-rely on AI-generated outputs~\cite{buccinca2021trust}, which is referred to as automation bias~\cite{romeo2026exploring}. Many issues in user trust are due to a lack of transparency~\cite{zerilli2022transparency, schmidt2020transparency}. According to Bach et al.~\cite{bach2024systematic}, user trust in AI-enabled systems can be influenced by technical and design features, as well as socio-ethical considerations and user characteristics. Regarding interventions, Bu{\c{c}}inca et al.~\cite{buccinca2021trust} found cognitive forcing to be effective in reducing overreliance on AI suggestions, though there was a tradeoff between the effectiveness and user acceptability as more effective interventions required more cognitive effort.

Regarding user control and agency, Zhang et al.~\cite{zhang2025exploring} mapped agency patterns with control mechanisms for different human-AI co-creation contexts. Moruzzi and Margarido~\cite{moruzzi2024user} proposed a user-centered co-creativity framework to modulate agency distribution between human users and the AI system based on a list of dimensions, which suggests how granular control mechanisms can help preserve human agency in mixed-initiative systems.

Safety guardrails should also be in place for human-AI collaborative systems to prevent erroneous system output and intentional unethical requests from the user~\cite{shamsujjoha2025swiss}. Gurita et al.~\cite{gurita2025breaking} deliberately instructed AI design tools to create an inaccessible user interface and discussed implications on systematically breaking AI-powered design tools to build better safety guardrails.

Based on these challenges in human-AI collaboration, Amershi et al.~\cite{amershi2019guidelines} proposed a set of 18 guidelines for designing human-AI interactive systems and emphasized the importance of transparency, error recovery, user control, and granular feedback. In line with these works, we study how confabulations in LLM-assisted immersive scene editing systems affect human-AI collaboration in terms of user trust and agency. We investigate whether confabulations in immersive scenes pose unique challenges to human-AI collaboration and use these findings to inform the design of confabulation mitigation strategies explicitly for immersive contexts.

\section{Methodology}

In this section, we present the method for implementing an LLM-assisted immersive scene editing workflow and details of the design of three scenes and three different types of tasks to provide a testbed to observe and study confabulations. This study is approved by the research ethics committee at
the University of Cambridge.

\subsection{Apparatus}

To collect interaction data to study different types of confabulation, we designed an LLM-assisted scene editing workflow in VR. This workflow follows existing works~\cite{yang2025llplace, zhao2026scenerevis, chen2025llmer, feng2023layoutgpt} to instruct the LLM to process the user natural language input and current scene information in JSON format produce a natural language response and optionally, the updated scene information in JSON format to make the edit.

\begin{figure*}[h!]
    \centering
    \includegraphics[width=0.8\linewidth]{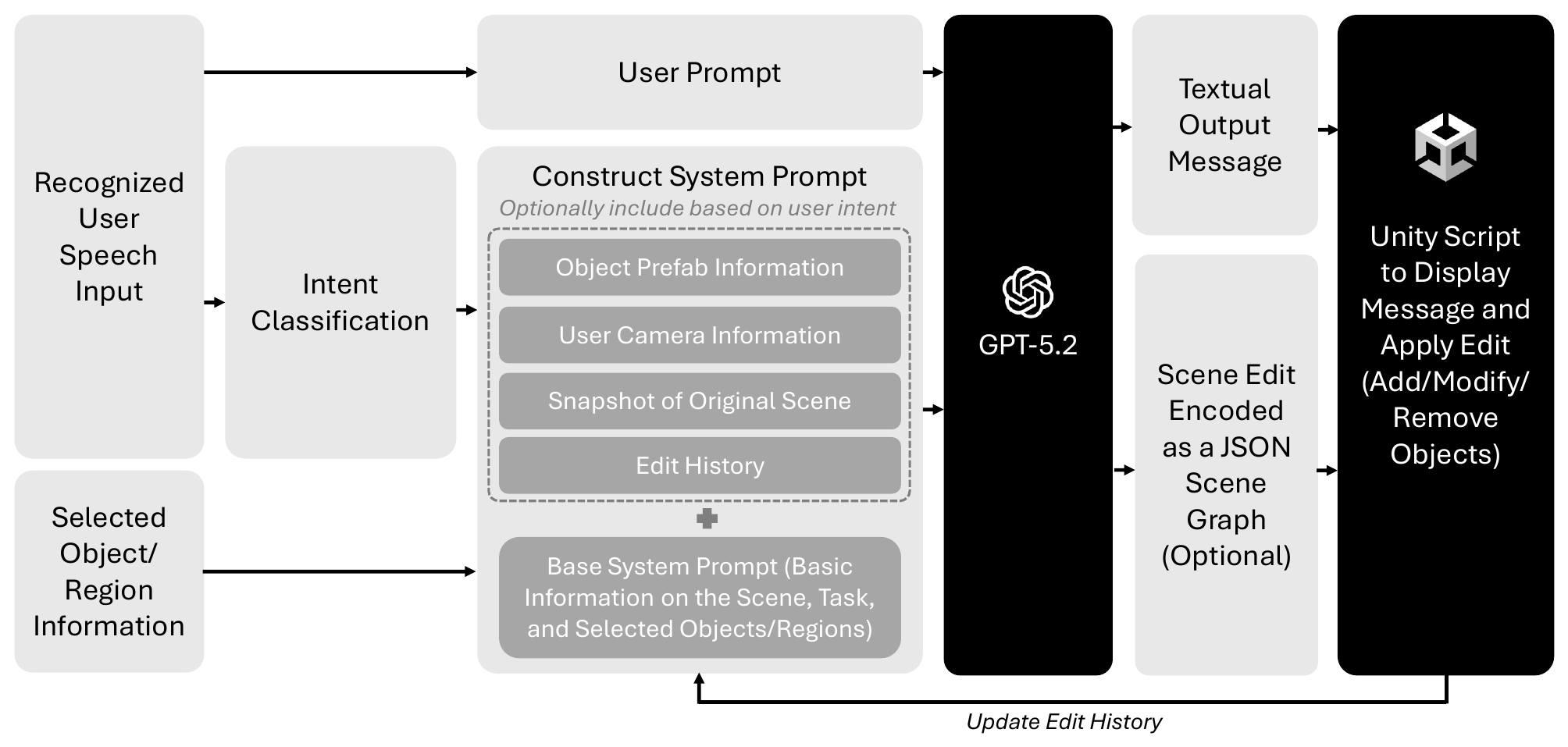}
    \caption{Workflow of the LLM-assisted immersive scene editing system adopted in the study.}
    \label{fig:system-workflow}
\end{figure*}

\Cref{fig:system-workflow} presents the workflow of the LLM-assisted scene editing system. The system takes the user speech and selected objects/regions as input. Speech is converted to text via a continuous speech recognizer. Subsequently, a rule-based classifier determines the user's intent by identifying regex patterns in the user speech input. Based on the user's intent, the system prompt is constructed. This includes a base system prompt containing basic information about the scene, the task, and selected objects/regions as well as optional information on editable object prefabs, the current user camera position and orientation, snapshot of the original scene in JSON scene graph format, and information on edit history (including object ID, its position and material attributes, and whether it was removed or added/modified). Subsequently, both the system prompt and user input are sent to the GPT-5.2 model by OpenAI, and the model outputs a textual message and optionally generates a JSON scene graph of edited objects in the scene. The message and scene graph are then processed by scripts in Unity to display the textual message on an interface in front of the user and apply the edit in the VR scene. If an edit is made, the edit history is updated.

This design was chosen to reflect the widely-adopted `JSON-in, JSON-out' LLM-assisted editing workflow in existing literature~\cite{yang2025llplace, zhao2026scenerevis, chen2025llmer, feng2023layoutgpt}. It also provides the flexibility for the system to answer to user queries and update the scene only if the instruction is clear, or ask for clarification otherwise. An additional intent classification step was added to ensure that only the most relevant information is passed to the LLM to improve efficiency and prevent information overload.

\begin{figure}[h!]
    \centering
    \includegraphics[height=5cm]{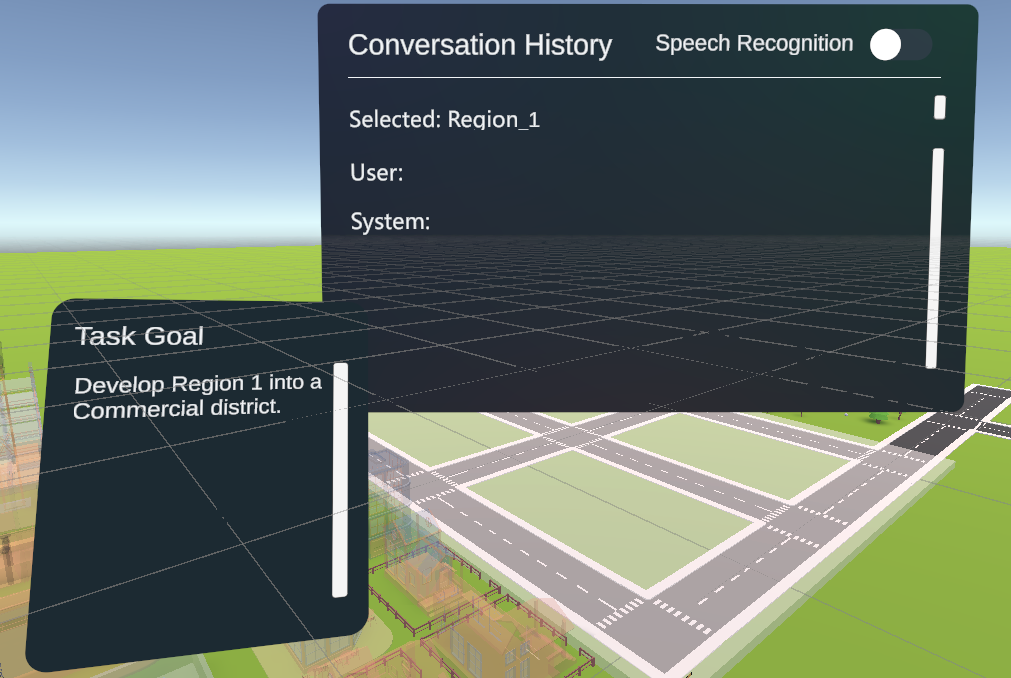}
    \includegraphics[height=5cm]{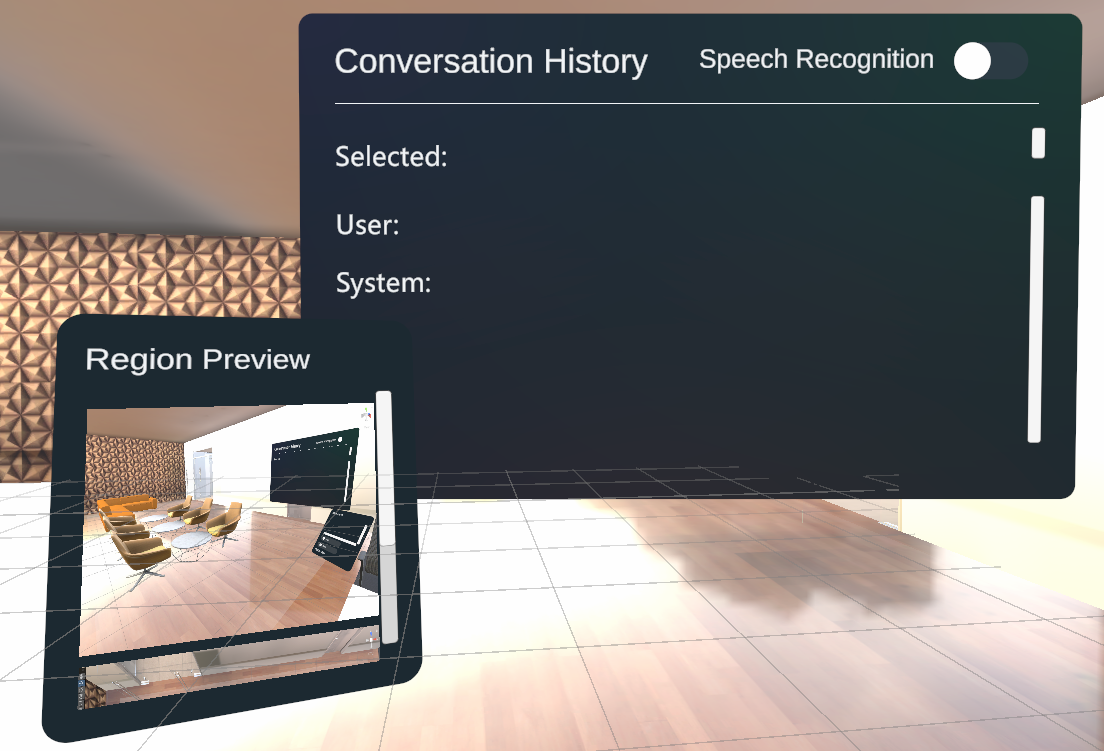}
    \caption{User interface of the LLM-assisted immersive scene editing system with the main panel containing selected objects and conversation history and the side panel containing textual task instructions (left) or image previews (right) of the target scene.}
    \label{fig:ui}
\end{figure}

The VR scene is implemented using Unity 3D (version 2022.3.15f1). The system user interface (UI) is implemented based on the Meta Interaction SDK~\cite{Meta_2025}. Examples of the UI in the office, loft, and city scenes are shown in \Cref{fig:teaser}. The main panel contains a continuous speech recognition toggle button, a list of currently selected objects, and the conversation history between the user and the system. An additional side panel with task instructions (\Cref{fig:ui} left) or image previews of the target scene appearance (\Cref{fig:ui} right) is provided to the left of the main panel. The UI allows users to toggle buttons, select/deselect objects in the scene, and slide text fields up or down to reveal more content.

Regarding hardware devices, participants wore an Oculus Quest 2 headset and interacted with scene elements via the left or right controller. The headset is connected to a Desktop device (Intel Core i7-6800K CPU, 64GB RAM, and NVIDIA TITAN X (Pascal) 12 GB graphics card) via an Oculus Quest Link cable.


\subsection{Participants}

The study involved 24 participants (12 male, 12 female), aged between 20 to 52 ($M=30.3, SD=8.0$). The participant sample shows a mix of users with different levels of expertise in VR, speech-recognition systems, and 3D design. 25\% of participants reported being experienced or very experienced with head-mounted VR, while 41.7\% reported being inexperienced or very inexperienced. Similarly, 29.2\% participants reported being experienced or very experienced with speech recognition systems, while 50\% reported being inexperienced or very inexperienced. For 3D scenes like CAD and video games, participants reported more experience, with 50\% reported being experienced or very experienced and 41.7\% reported being inexperienced or very inexperienced. 50\% of participants reported being native English speakers and the other 50\% self-identified as non-native English speakers. Only one out of all 24 participants reported accessibility needs (the participant self-identified as being red-green color blind), though the participant was able to distinguish between colors of objects in the scene.

\subsection{Task and Scene Design}

To study different types of confabulation in LLM-assisted immersive scene editing, we carefully designed the scenes and tasks to represent as many use cases of such systems as possible. The tasks are designed to reflect single- and multi-object editing tasks (e.g., adding single or multiple objects in the scene with a specified position and orientation, or modifying properties of single or multiple objects which already exist in the scene) as well as high-level design tasks such as stylistic edits (changing all objects in the scene to match a target style) or functional design (designing the space to fulfill a high-level functional requirement) of the entire scene. The latter two task types where high-level edits are involved for complex task-solving are exactly the cases where speech-driven interfaces show greater promise as compared to direct manipulation interfaces~\cite{shneiderman1997direct, feng2024large}. Therefore, two dedicated scenes are designed to study confabulations in stylistic edit tasks and functional design tasks respectively.

Three scenes were designed to reflect different LLM-assisted editing task contexts and accommodate the low-level single-/multi-object editing tasks and high-level stylistic edit and functional design tasks. These are the \textbf{office}, \textbf{loft}, and \textbf{city} scenes. Example screenshots of the three scenes are provided in \Cref{fig:teaser}. Each scene represented only one task type, with each task type containing three variants. For the office scene, participants are instructed to edit a specified region to match the target appearance (three target appearances were involved in total for the office scene, which constitute the three task variants) in the image in the side panel (see \Cref{fig:ui} bottom). For the loft scene, participants are asked to edit the materials of objects in the living room to match a given target style (three task variants with target styles specified in text, e.g., dark monochrome/warm cozy/sea blue style). For the city scene, participants are asked to add buildings in the specified region to form a residential/commercial/industrial district, which correspond to the three task variants for the city scene. 

The office and loft scene reflect small-scale indoor edit scenarios, while the city scene reflects large-scale outdoor editing scenarios. The office scene focuses on the placement edits of new and existing objects, while the loft scene targets high-level stylistic edits of the entire scene, and the city scene focuses on high-level edits of large city regions to satisfy a functional requirement. Together, the choice of these three scenes span scale (small to large), complexity (low to high), different edit types (object position/orientation edits and object appearance edits) and different task types (fine-grained object-level edits and high-level scene edits).

\begin{table*}[h!]
\centering
\caption{Distribution of different types of confabulation across the office, loft, and city scenes. The percentage of a confabulation subtype in a scene with respect to total confabulation turns in that scene, as well as with respect to total interaction turns in that scene are shown in brackets in each cell. Number of occurrences do not necessarily sum up to the total values as each interaction turn can be assigned with multiple types and subtypes.}
\label{tab:confabulation-type-distribution}
\renewcommand{\arraystretch}{1.4}
\resizebox{\textwidth}{!}{
\begin{tabularx}{1.337\textwidth}{
>{\raggedright\arraybackslash}p{0.23\textwidth}
>{\raggedright\arraybackslash}p{0.6\textwidth}
p{0.13\textwidth}p{0.12\textwidth}p{0.14\textwidth}}
\hline
\multirow{2}{*}{\textbf{Confabulation   Types}} &
  \multicolumn{1}{c}{\multirow{2}{*}{\textbf{Subtypes}}} &
  \multicolumn{3}{c}{\textbf{Number of Occurrences}} \\ \cline{3-5} 
 &
  \multicolumn{1}{c}{} &
  \textbf{Office} &
  \textbf{Loft} &
  \textbf{City} \\ \hline
\multirow{5}{=}{\textbf{Spatial Placement Error (SPE)}}
&
  SPE-OUT: Out-of-region placement &
  33 (22\%, 5\%) &
  0 (0\%, 0\%) &
  115 (47\%, 19\%) \\
 &
  SPE-ROAD: Road/boundary overlap &
  0 (0\%, 0\%) &
  0 (0\%, 0\%) &
  117 (48\%, 19\%) \\
 &
  SPE-COL: Collision between objects &
  51 (34\%, 8\%) &
  0 (0\%, 0\%) &
  59 (24\%, 10\%) \\
 &
  SPE-DIS: Error in displacement   direction/magnitude &
  1 (1\%, 0\%) &
  0 (0\%, 0\%) &
  7 (3\%, 1\%) \\ \cline{2-5}
 &
  \textbf{SPE TOTAL} &
  \textbf{76 (51\%, 12\%)} &
  \textbf{0 (0\%, 0\%)} &
  \textbf{223 (91\%, 37\%)} \\ \hline
\multirow{3}{=}{{\textbf{Reference Frame Ambiguity (RFA)}}} &
  RFA-DIR: Directional inversion &
  25 (17\%, 4\%) &
  0 (0\%, 0\%) &
  5 (2\%, 1\%) \\
 &
  RFA-LMK: Landmark reference failure &
  3 (2\%, 0\%) &
  0 (0\%, 0\%) &
  0 (0\%, 0\%) \\ \cline{2-5}
 &
  \textbf{RFA TOTAL} &
  \textbf{28 (19\%, 5\%)} &
  \textbf{0 (0\%, 0\%)} &
  \textbf{5 (2\%, 1\%)} \\ \hline
\multirow{3}{=}{{\textbf{False Scene Claim (FSC)}}}
&
  FSC-DONE: False `already done' claim &
  7 (5\%, 1\%) &
  3 (7\%, 1\%) &
  18 (7\%, 3\%) \\
 &
  FSC-MISMATCH: Description-action mismatch &
  0 (0\%, 0\%) &
  1 (2\%, 0\%) &
  1 (0\%, 0\%) \\ \cline{2-5}
 &
  \textbf{FSC TOTAL} &
  \textbf{7 (5\%, 1\%)} &
  \textbf{4 (9\%, 1\%)} &
  \textbf{19 (8\%, 3\%)} \\ \hline
\multirow{4}{=}{{\textbf{Command Misinterpretation without Disclosure (CMD)}}} &
  CMD-TOKEN: Misinterpretation of an unknown token without disclosure &
  0 (0\%, 0\%) &
  20 (45\%, 5\%) &
  0 (0\%, 0\%) \\
 &
  CMD-INCOMPLETE: Acting on an incomplete   command without disclosure &
  3 (2\%, 0\%) &
  1 (2\%, 0\%) &
  4 (2\%, 1\%) \\
 &
  CMD-AMB: Acting upon an ambiguous   reference without disclosure &
  3 (2\%, 0\%) &
  2 (5\%, 0\%) &
  0 (0\%, 0\%) \\ \cline{2-5}
 &
  \textbf{CMD TOTAL} &
  \textbf{6 (4\%, 1\%)} &
  \textbf{23 (52\%, 6\%)} &
  \textbf{4 (2\%, 1\%)} \\ \hline
\multirow{4}{=}{{\textbf{Edit Scope Error (ESE)}}} &
  ESE-OVER: Over-scope edit error &
  6 (4\%, 1\%) &
  1 (2\%, 0\%) &
  0 (0\%, 0\%) \\
 &
  ESE-UNDER: Under-scope edit error &
  3 (2\%, 0\%) &
  2 (5\%, 0\%) &
  0 (0\%, 0\%) \\
 &
  ESE-WRONG: Edit of wrong object &
  2 (1\%, 0\%) &
  1 (2\%, 0\%) &
  0 (0\%, 0\%) \\ \cline{2-5}
 &
  \textbf{ESE TOTAL} &
  \textbf{11 (7\%, 2\%)} &
  \textbf{4 (9\%, 1\%)} &
  \textbf{0 (0\%, 0\%)} \\ \hline
\multirow{4}{=}{{\textbf{Orientation Error (ORI)}}} &
  ORI-FACE: Object facing direction error &
  12 (8\%, 2\%) &
  0 (0\%, 0\%) &
  0 (0\%, 0\%) \\
 &
  ORI-GROUP: Group rotation error &
  1 (1\%, 0\%) &
  0 (0\%, 0\%) &
  0 (0\%, 0\%) \\
 &
  ORI-UNINSTR: Uninstructed rotation &
  0 (0\%, 0\%) &
  0 (0\%, 0\%) &
  1 (0\%, 0\%) \\ \cline{2-5}
 &
  \textbf{ORI TOTAL} &
  \textbf{13 (9\%, 2\%)} &
  \textbf{0 (0\%, 0\%)} &
  \textbf{1 (0\%, 0\%)} \\ \hline
\multirow{4}{=}{{\textbf{Material/Style Error (MSE)}}}
&
  MSE-STYLE: Style-inconsistent edit &
  0 (0\%, 0\%) &
  4 (9\%, 1\%) &
  0 (0\%, 0\%) \\
 &
  MSE-COLOR: Color interpretation error &
  0 (0\%, 0\%) &
  6 (14\%, 1\%) &
  0 (0\%, 0\%) \\
 &
  MSE-UNINSTR: Uninstructed color/material   edit &
  0 (0\%, 0\%) &
  1 (2\%, 0\%) &
  0 (0\%, 0\%) \\ \cline{2-5}
 &
  \textbf{MSE TOTAL} &
  \textbf{0 (0\%, 0\%)} &
  \textbf{11 (25\%, 3\%)} &
  \textbf{0 (0\%, 0\%)} \\ \hline
\multirow{2}{=}{{\textbf{Missing Information without Disclosure (MID)}}} &
  MID-CONTEXT: Missing context information without disclosure &
  5 (3\%, 1\%) &
  2 (5\%, 0\%) &
  3 (1\%, 0\%) \\ \cline{2-5}
 &
  \textbf{MID TOTAL} &
  \textbf{5 (3\%, 1\%)} &
  \textbf{2 (5\%, 0\%)} &
  \textbf{3 (1\%, 0\%)} \\ \hline
\multirow{4}{=}{{\textbf{Action Execution Failure (AEF)}}}
&
  AEF-EDIT: Claimed edit with failed execution &
  4 (3\%, 1\%) &
  1 (2\%, 0\%) &
  0 (0\%, 0\%) \\
 &
  AEF-DELETE: Claimed deletion with failed   execution &
  2 (1\%, 0\%) &
  0 (0\%, 0\%) &
  0 (0\%, 0\%) \\
 &
  AEF-UNDO: Claimed undo with failed   execution &
  2 (1\%, 0\%) &
  0 (0\%, 0\%) &
  0 (0\%, 0\%) \\ \cline{2-5}
 &
  \textbf{AEF TOTAL} &
  \textbf{8 (5\%, 1\%)} &
  \textbf{1 (2\%, 0\%)} &
  \textbf{0 (0\%, 0\%)} \\ \hline
\multicolumn{2}{c}{\textbf{Total Interaction Turns with Confabulation}} &
  \textbf{149 (-, 24\%)} &
  \textbf{44 (-, 11\%)} &
  \textbf{244 (-, 41\%)} \\ \hline
\multicolumn{2}{c}{\textbf{Total Interaction Turns}} &
  \textbf{609} &
  \textbf{402} &
  \textbf{601} \\ \hline
\end{tabularx}
}
\end{table*}

\subsection{Procedure}

Participants are greeted and invited to read an information sheet and sign a consent form at the beginning of the study. Subsequently, participants filled out a demographic survey and received instructions for the study. This involved explaining the study timeline, providing an overview of different scenes and tasks for each scene, providing basic instructions on how to use the system hardware and software, as well as showing a demo video of issuing a sample command in each scene. Participants were not briefed on the specific instructions that they could issue and had the flexibility to phrase their own commands and choose their own approach to complete the specified task.

Subsequently, participants completed one set of trials for each of the three scenes. For each scene, participants completed a practice trial (maximum 1.5 minutes), trial 1 (maximum 4 minutes), and trial 2 (maximum 4 minutes), followed by post-experience questions for that scene. There were three task variants for each of the three scenes, and each task variant involved a different task goal (for example a different set of target images for the office scene, different target styles for the loft scene, and different functional requirements for the city scene). The order of different task variants and the order of the three scenes for each participant were counterbalanced across all participants. Each trial ended when participants were satisfied with the edit or if the maximum time limit was reached. As participants attempted the tasks, the system logged timestamped conversation history, timestamped edit history with scene graph snapshots whenever an edit was made, select/deselect actions, and logs from the console. Post-experience questions for each individual scene involved questions on observed confabulations (including the type, frequency level of disruption, and a short description of the confabulation), questions on user trust, confidence, and satisfaction, as well as questions from the unweighted NASA-TLX~\cite{hart1988development} questionnaire.

After participants completed all trials in all three scenes, they were invited to fill out a questionnaire regarding the overall experience. This included the System Usability Scale (SUS)~\cite{brooke1996sus} and questions which involved ranking the overall frequency and severity of confabulations, as well as the difficulty of the three task types associated with each of the three scenes. Participants also completed open-ended questions to comment on the types of confabulation that were most problematic, surprising and confusing confabulation behavior, improvements for future systems, and other open comments. In total, the study session lasted 60 to 75 minutes for each participant. Participants were thanked for their time and remunerated at the end of the study. 


\section{Results}

\subsection{A Taxonomy of Confabulations Based on Empirical Study Log Data (RQ1)}\label{sec:taxonomy}

Based on the interaction log data of all 24 participants, we construct a taxonomy of different types of confabulation in LLM-assisted immersive scene editing. Through the study, a total of \textbf{1612 interaction turns} were logged. For each interaction turn, the user speech input (recognized and recorded as text), system textual output, as well as information on selected objects, and the changes applied to the scene during that step (if any) are recorded. Each interaction turn was manually reviewed to compare the user input, system output, and a replay and visual inspection of the scene edit to identify whether confabulation was present.

Overall, the log data yielded \textbf{437 confabulation turns} (27.1\% of total interaction turns). Confabulation turns were then analyzed with an inductive open coding approach~\cite{saldana2021coding} to build a bottom-up taxonomy, which identified nine main confabulation types and 24 subtypes. Each interaction turn could be labeled with one or multiple confabulation types/subtypes. A complete summary of these confabulation types, subtypes, and their number of occurrences across the office, loft, and city scenes is shown in \Cref{tab:confabulation-type-distribution}. As the total number of interaction turns and confabulation turns is different across the three scenes, we also report the percentage of confabulation subtypes with respect to the total number of confabulation turns and the total number of interaction turns in that scene to indicate the prevalence of that confabulation subtype in the scene.

The first cluster of confabulation types includes failures related to the \textbf{spatial interpretation of the scene} (SPE, RFA, ORI). \textbf{Spatial placement errors (SPE)} refer to cases where the LLM places objects at an incorrect location (such as outside the region boundary or at positions overlapping with roads or other objects), despite being given explicit spatial constraints. This confabulation type is the most prevalent across all confabulation types (299 confabulation turns among the total 437 confabulation turns), accounting for 91\% of the total confabulation turns in the city scene and 51\% of confabulation turns in the office scene. Three subtypes dominated, including out-of-region placements (SPE-OUT), road/boundary overlaps (SPE-ROAD), and inter-object collisions (SPE-COL). The prevalence of these subtypes suggest a systematic inability of the system to respect spatially constrained regions during object placement. SPE-COL was the dominant subtype in the office scene (51 out of 76 SPE instances), which suggested the lack of collision awareness during furniture placement. SPE-DIS ($n=9$) captured cases when the move direction was correct but the magnitude is grossly wrong, resulting in an overshoot or no visible changes. SPE affected every one of the 24 participants which further highlights its prevalence in high-level functional design tasks.

\textbf{Reference Frame Ambiguity (RFA)} refers to cases where the LLM interprets user egocentric descriptions (front, back, forward, behind, etc.) using global coordinates, or vice versa. These account for 33 instances in the office and city scene. A dominant subtype is directional inversion (RFA-DIR, $n=30$), where the system used global coordinates rather than the user's viewing direction to interpret directions, causing objects to be displaced in a direction opposite to the user's intent. Landmark reference failures (RFA-LM, $n=3$) occurred when users named visible elements (e.g., `the back wall', `the glass door') as spatial anchors, but the LLM lacked positional data for these elements and defaulted to the global coordinate extremes.
RFA was unevenly distributed across participants. P5 and P22 each generated 4 RFA-DIR instances in the office scene, which suggests that RFA could be partly driven by how individuals phrased spatial instructions.

\begin{figure*}[h!]
    \centering
    \includegraphics[width=0.28\linewidth]{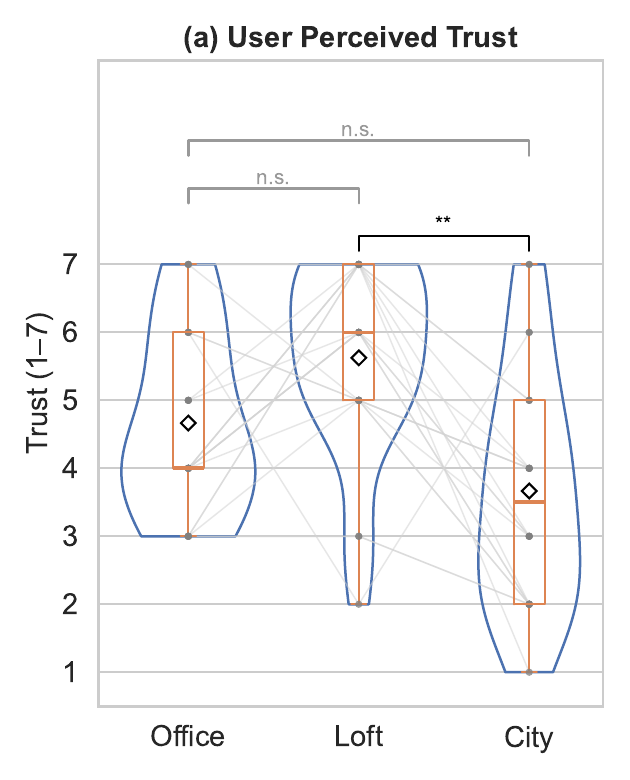}
    \includegraphics[width=0.28\linewidth]{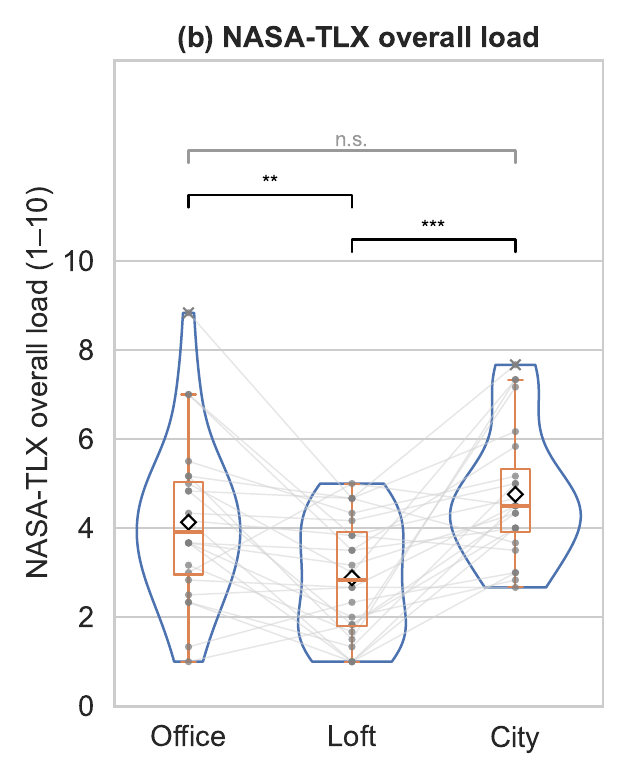}
    \includegraphics[width=0.28\linewidth]{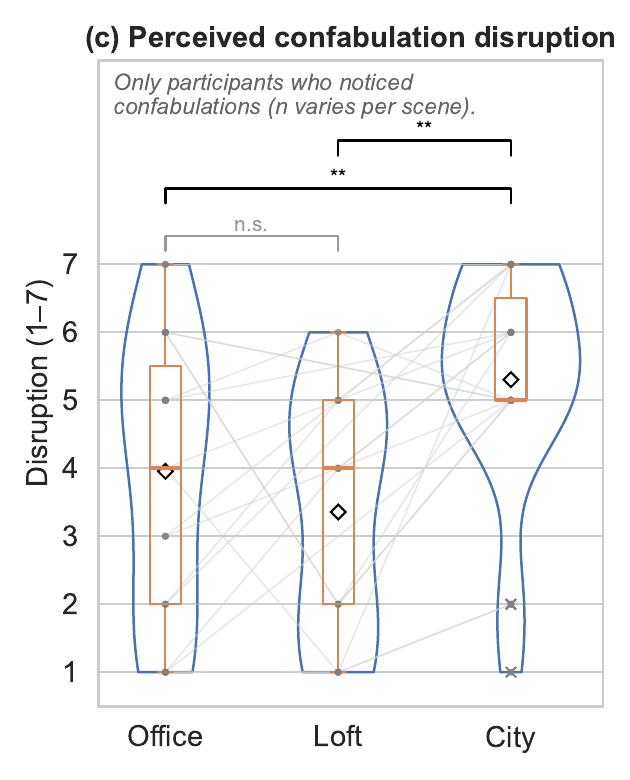}
    \caption{Violin plots of the user perceived trust ratings on a scale of 1 to 7 (left), unweighted NASA-TLX overall load from a scale of 1 to 10 (middle), and the perceived disruption of confabulations (when occurred) from a scale of 1 to 7 (right). Diamonds indicate the mean and asterisks indicate the level of significance in post-hoc tests: *$p<.05$, **$p<.01$, ***$p<.001$, n.s.: Not significant. Thin grey lines connect data from the same participant across different scenes.}
    \label{fig:trust-load-disruption}
\end{figure*}

\textbf{Orientation Errors (ORI)} refer to objects being rotated to orientations that misalign with the user's intent. 14 instances are recorded which dominated in the office scene ($n=13$). These included errors where objects fail to rotate to face each other when requested (ORI-FACE, $n=12$), errors in the orientation of individual objects when a group of objects were asked to rotate as a whole (ORI-GROUP, $n=1$), and rotations applied to objects when this is not instructed (ORI-UNINSTR, $n=1$).

The second cluster includes failures related to the \textbf{semantic interpretation of text information} (CMD, FSC, MID, MSE). \textbf{Command Misinterpretation without Disclosure (CMD)} refers to cases where the LLM receives an ambiguous of incomplete input but fails to ask for clarification or disclose its uncertainty and instead silently maps the input to its nearest plausible meaning. It is the dominant confabulation type in the Loft scene (23 out of 44 instances, 52.3\%). Its most prevalent subtype where specific tokens are misinterpreted (CMD-TOKEN, $n=20$) affected 12 out of 24 participants due to a specific vocabulary failure. The VR system uses an `\_st' suffix to store semitransparent material variants. The system did not have access to this information but consistently misinterpreted this as a stylistic variant and applied these materials without disclosing its lack of knowledge. CMD-INCOMPLETE ($n=8$) captured a similar failure mode where voice commands were cut off but the system continued to infer and produce an action. CMD-AMB ($n=5$) covered ambiguous object references and action verbs that required clarification but were instead silently interpreted.

\textbf{False Scene Claims (FSC)} refer to cases where the LLM made a wrong claim about the current state of the scene. These include 30 instances across the three scenes. The most dominant subtype is cases where the LLM incorrectly believed an action had been completed and refused to act (FSC-DONE, $n=28$). In this city scene, FSC-DONE typically followed cases where buildings are incorrectly placed outside the region. The system assumes the previous placement is correct, and refuses to move the building inside the region. Another subtype is
FSC-MISMATCH ($n=2$). Here, the system does not refuse to act (as is the case in FSC-DONE) but instead claims to make an edit which is inconsistent with its actual action. If the user relied on only the textual output message without verifying the actual edit, the user would assume the edit had been correctly made.

\textbf{Missing Information Without Disclosure (MID)} refers to cases where the LLM proceeds with an edit that requires information it does not have access to. The only subtype recorded is MID-CONTEXT ($n=10$), which is present in all three scenes. Examples include cases where the LLM does not have access to information such as the position and bounding box size of objects already in the city region (as they are not selected, their information is not passed to the LLM), the color of background objects in the loft scene, and the size of different chairs in the office scene. The LLM lacked the above information without disclosure, and issued edits which resulted in collision with objects already in the region, incorrect color edits, and incorrect choice of chairs to edit respectively.

\textbf{Material/Style Error (MSE)} refers to cases where the LLM applies colors and/or materials to objects which are inconsistent with the color, material, or style specified by the user. They are exclusive to the loft scene ($n=11$), which reflects the unique demands of stylistic editing tasks. MSE-COLOR ($n=6$) arose when color descriptions were mapped to incorrect materials, such as mapping `lighter blue' to white and mapping `warm but darker' to grey. MSE-STYLE ($n=4$) captured style-inconsistent edits, such as using seafoam colors (a light blue-green hue) when a warm style was requested. One uninstructed color/material edit (MSE-UNINSTR) was also present. MSE and CMD together constitute 34 out of the total 44 confabulation instances (77.2\%) in the loft scene, which suggest how stylistic editing tasks are dominated by semantic-level rather than spatial-level failures.

The third cluster includes failures related to \textbf{execution of edit or undo actions} (ESE, AEF).
\textbf{Edit Scope Error (ESE)} refers to cases where the LLM edits an incorrect set of objects compared to what the user specified. Cases were concentrated in the office scene ($n=11$), followed by the loft scene ($n=4$). ESE-OVER ($n=7$) involved edits beyond the intended set of objects and commonly occurred when collective noun objects were referenced (commands like `rotate all chairs' caused prefab chairs to also be rotated). ESE-UNDER ($n=5$) captured cases where only a subset of the intended objects were edited, for example when only certain objects were edited when the user requested the style to be applied to `each item', or when only chairs are placed when chairs and a meeting table are requested together. ESE-WRONG ($n=3$) captured cases where the edit was applied to a different set of objects.

\textbf{Action Execution Failure (AEF)} refers to cases where the LLM claims to have performed an action (e.g., deletion/undo/duplication/edit) when the action is absent, partially executed, or results in an outcome inconsistent with the claimed action. 9 instances are recorded across the office ($n=8$) and loft scene ($n=1$). For example, P10 was told that the chair has been deleted, yet the chair remains in the scene. P13 was told that an undo was applied when the undo action was not completed. AEF and FSC are less frequent but arguably the most detrimental for user trust as the system provided a message that the correct edit was applied. Unless the user visually inspected the action outcomes or scene state, such confabulations will go unnoticed.

In summary, nine confabulation types grouped into three clusters were identified based on the interaction log data. Different task types revealed different dominant confabulation types. Functional high-level design tasks in constrained spatial grids (city scene) were dominated by placement failures (SPE). Object-level positional and rotational edits (office scene) produced a broad multi-type profile which included boundary interpretation errors which caused collisions between objects (SPE-COL) and between objects and boundaries (SPE-OUT), as well as directional and orientation errors (RFA, ORI). Stylistic editing tasks (loft scene) produced a distinct non-spatial failure mode, with dominant types spanning vocabulary (CMD) and color/style (MSE) interpretation failures. Crucially, silent failures where the LLM produces a plausible and confident response while implementing the edit incorrectly (CMD, MID, FSC, and RFA-DIR) are present across all three task types and poses a core design challenge for LLM-assisted scene editing systems. In the subsequent subsections, we will further investigate the effects of confabulations on user trust and the perception-reality gap of confabulations to present more empirical insights.

\begin{figure*}[h!]
    \centering
    \includegraphics[height=5.425cm]{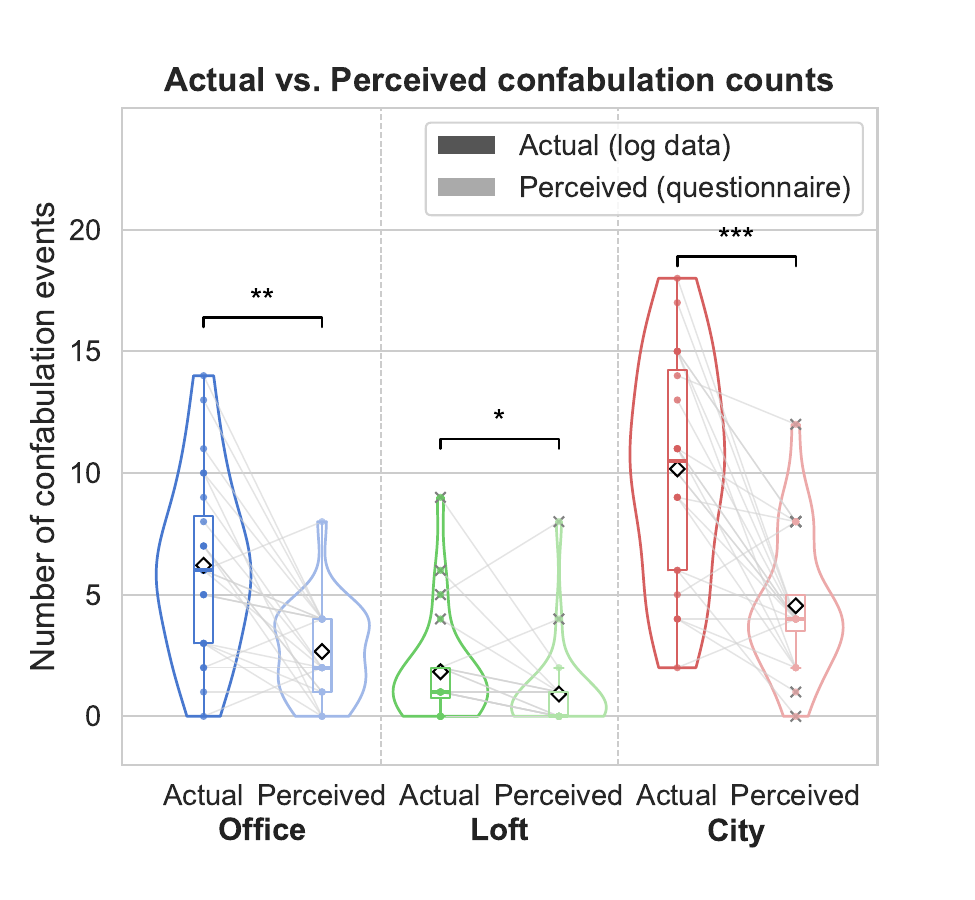}
    \includegraphics[height=5.425cm]{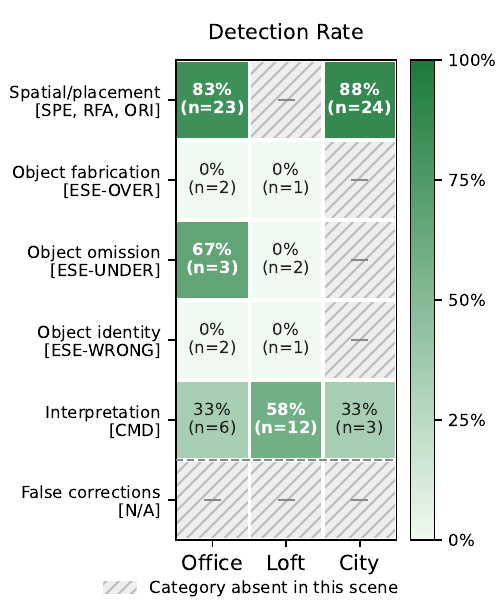}
    \includegraphics[height=5.425cm]{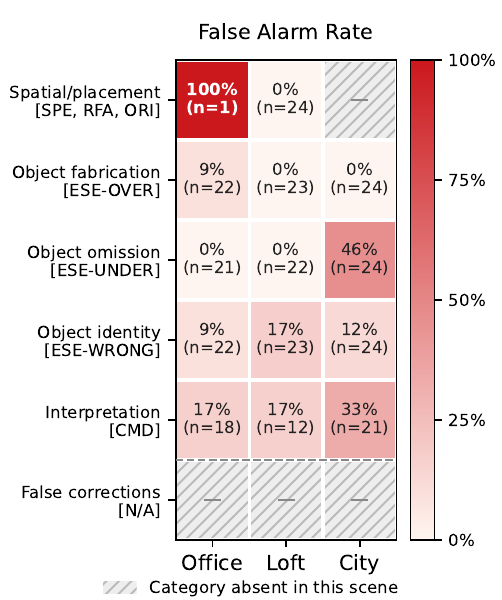}
    \caption{Actual and perceived confabulation counts in both trials of all participants in the office, loft, and city scenes (left) with diamonds indicating the mean and asterisks indicating the level of significance in post-hoc tests (*$p<.05$, **$p<.01$, ***$p<.001$) and thin grey lines connecting data points from the same participant, and the detection (middle) and false alarm (right) rates for each confabulation type and scene. $n$ indicates group size, and hatched cells indicate that the category was absent in the scene or the questionnaire confabulation type could not be mapped to a confabulation type in \Cref{sec:taxonomy}.}
    \label{fig:actual-perceived}
\end{figure*}

\subsection{Confabulation Disruptions and their Impact on User Trust and Workload (RQ2)}\label{sec:trust-load-disrupt}

In this subsection, we present ratings of user perceived trust in the system, NASA-TLX overall load, and perceived level of disruption when confabulations occurred to study how confabulations affected users across the task types presented in the three different scenes. Friedman tests and post-hoc pairwise Wilcoxon signed-rank tests with Bonferroni correction were used to determine statistically significant differences and report effect sizes (rank-biserial correlation $r$). \Cref{fig:trust-load-disruption} presents a violin plot of the three types of ratings of all participants across all three scenes, with thin grey lines connecting data from the same participant.

\paragraph{Perceived Trust.} Trust ratings were obtained by asking users to rate from 1-Strongly Disagree to 7-Strongly Agree on the statement `I trusted the system to implement my edits correctly'. Trust ratings were significantly different across the three scenes (Friedman $\chi^2=21.60, p<.001$) and post-hoc tests revealed that the trust rating in the city scene ($M=3.67, SD=1.71$) was significantly lower ($W=193.5, p<.001, p_{bonf}=.003, r=.67$) than that of the loft scene ($M=5.62, SD=1.41$). Pairwise comparisons between the office ($M=4.67, SD=1.34$) and loft scene ($p_{bonf}=.074, r=.46$) and the office and city scene ($p_{bonf}=.052, r=.49$) revealed no significant differences. The overall pattern suggests that trust was lowest in the city scene (representing high-level functional design tasks in scenes with complex spatial constraints) where confabulation was most prevalent (\Cref{sec:taxonomy}).

\paragraph{Overall Load.} Overall load was calculated by averaging across the unweighted NASA-TLX subscales from a Likert scale of 1-Very Low Workload to 10-Very High Workload. Friedman tests revealed a significant difference in overall load across the three scenes ($\chi^2=17.62, p<.001$), and post-hoc tests showed that the overall load was significantly lower in the loft scene ($M=2.88, SD=1.32$) compared to both the office scene ($M=4.13, SD=1.82, W=259.5, p<.01, p_{bonf}=.006, r=.64$) and city scene ($M=4.76, SD=1.48, W=12.5, p<.001, p_{bonf}<.001, r=.78$) respectively. Significant differences were not found between the overall load of the office and city scene ($W=98.5, p_{bonf}=.435, r=.30$). These results suggest that confabulations in object-placement and functional design tasks posed a comparable higher load compared with stylistic editing tasks.

\paragraph{Confabulation Disruption.} Disruption ratings were only collected from participants who noticed confabulations in a certain scene ($n_{office}=23, n_{loft}=14, n_{city}=23$), where participants are asked to rate from a scale of 1-Strongly Disagree to 7-Strongly Agree on the statement `When a hallucination occurred, it disrupted my task'. Among those who noticed confabulations, Friedman tests indicated a significant difference in disruption ratings across scenes ($\chi^2=12.41, p<.01, n=14$ for complete cases across all scenes). Post-hoc tests revealed that the disruption rating was significantly higher in the city scene ($M=5.30, SD=1.66$) compared to the office ($M=3.96, SD=2.01, W=18.0, p<.01, p_{bonf}=.010, r=.63, n=22$) and loft scenes ($M=3.36, SD=1.78, W=3.0, p<.01, p_{bonf}<.01, r=.79, n=14$) respectively. Disruption scores did not significantly differ between the office and loft scenes ($W=23.0, p_{bonf}=1.0, n=14$).

The three measures demonstrate how the loft scene with stylistic design tasks had the highest level of perceived trust, lowest load level, and lowest disruption when confabulations occurred, while the city scene with functional design tasks and complex spatial constraints yielding the worst outcomes in trust, load, and disruption. This order aligns with findings from the confabulation taxonomy in \Cref{sec:taxonomy}. Color and material confabulations in stylistic edits were less prevalent and easy to correct, while spatial confabulations were persistent and had a negative impact in all three measures.

\subsection{The Gap between Perceived and Actual Confabulations (RQ3)}\label{sec:gap}

In this subsection, we compare perceived confabulations (based on post-experience questionnaire data\footnote{The questionnaire involved pre-defined confabulation types to probe perceived confabulation types, which was before the study when actual confabulation types were not known. Confabulation types identified in \Cref{sec:taxonomy} were constructed using a bottom-up approach, and are mapped to predefined categories in the questionnaire (see \Cref{fig:actual-perceived} middle and right). FSC, AEF, and MSE did not correspond to questionnaire items, which resulted in a structural blind spot. False corrections in the questionnaire did not correspond to any confabulation types and are not applicable in the heatmap.}) against actual confabulations (based on analyzed log data in \Cref{sec:taxonomy}) to identify gaps and suggest cases where silent confabulations are most prevalent, followed by a brief discussion on factors which lead to this perception-reality gap. Through this subsection, we extend the contribution of our work beyond this particular system workflow to provide general implications for the integration of intelligent systems in immersive space where confabulations could potentially occur.

\paragraph{Systematic underestimation of confabulation frequency.}
A comparison between the number of confabulation occurrences in \Cref{sec:taxonomy} against a reported estimation of the number of confabulations across the two trials for each scene from each participant revealed a systematic underestimation of confabulation frequency across all three scenes (\Cref{fig:actual-perceived} left). Pairwise Wilcoxon signed-rank tests with Holm correction are used to test whether perceived confabulation counts would be lower than actual counts within each scene. Effect sizes are reported as rank-biserial correlation. The results suggest that the perceived counts (Office $Mdn=2.0$, Loft $Mdn=0.0$, City $Mdn=4.0$) are significantly lower than the actual counts (Office $Mdn=6.0$, Loft $Mdn=1.0$, City $Mdn=10.5$) in all three scenes (Office $W=22.5, p_{adj}<.01, r=.69$, Loft $W=22.5, p_{adj}<.05, r=.44$, City $W=10.5, p_{adj}<.001, r=.79$).

Spearman rank correlations are reported to assess whether participants who experience more confabulations also perceive more. Correlations between the perceived and actual counts for the office ($\rho=.15, p=.47$) and city ($\rho=.30, p=.16$) scenes were not significant, suggesting that the number of perceived counts in the two scenes are unrelated to the number of confabulations actually encountered by the participant. In the loft scene ($\rho=.71, p<.001$) with fewer actual confabulations ($Mdn=1$), users do track confabulations well at an individual level. A Friedman test confirmed that both actual ($\chi^2=32,8, p<.001$) and perceived counts ($\chi^2=29.6, p<.001$) differed significantly at the scene level. Post-hoc pairwise Wilcoxon tests with Holm correction showed that all three scene pairs were significantly different for the actual ($p_{adj}<.001$ for all three pairwise comparisons) and perceived counts ($p_{adj}<.01$ for all three pairwise comparisons) respectively.

Together, these results point to an interesting interpretation of the perception-reality gap. At a low confabulation load in the loft scene, users do track confabulations well at the individual level. However, under high confabulation load in the office and city scenes, the ability to perceive confabulations saturates and breaks down, causing perceived confabulations to fall significantly short of actual confabulations.

\paragraph{Type-level Detection and False Alarms.}
\Cref{fig:actual-perceived} also shows detection rates (proportion of participants who experienced a confabulation type and identified it in the questionnaire, middle subfigure) and false alarm rates (proportion who reported a type they did not experience, right subfigure) of different confabulation types in the three scenes. Hatched cells indicate that the category was absent from the scene ($n_{actual}=0$) or if the questionnaire confabulation type could not cleanly map to the taxonomy in \Cref{sec:taxonomy}. Formal cell-level tests were not conducted due to the limited number of affected participants for many cells and descriptive results are reported instead.

Regarding detection rate, spatial and placement errors (SPE, RFA, ORI) were consistently identified: 19 out of 23 affected participants (83\%) in the office scene and 21 out of 24 affected (88\%) in the city scene reported this confabulation, suggesting that spatial violations can be perceived by most participants in the immersive scene. Interpretation errors were reported by 7 out of 12 affected participants (58\%) in the loft scene and 2 out of 6 (33\%) in the office scene, which also shows a moderate detection accuracy. The remaining confabulation subtypes each affected at most 3 participants in each scene, which limits how stable observations can be made from the data.

Regarding false alarms, \Cref{fig:actual-perceived} (right) shows the endorsement rate of participants who did not experience the confabulation type. The most substantial finding is in the city scene, where 11 out of 24 participants (46\%) reported object omission when this error was completely absent. Further inspection of the log data suggested that 6 of these 11 participants experienced FSC-DONE confabulations, where the system falsely claimed that an action had been completed when it was not. Notes taken during the study also showed that some participants did not find objects which were misplaced outside the region or were misplaced among the prefab objects and were unnoticed by the user, which suggests a possible explanation of why users believed that object omission (where requested objects were not created) occurred. Interpretation errors also attracted some false alarms in the city (7/21, 33\%) and office scene (3/18, 17\%), which suggests that some salient failures are retrospectively attributed to intent misinterpretation. Object identity errors were falsely reported in the loft scene (4/23, 17\%), and object fabrication was falsely reported in the office scene (2/22, 9\%). The loft scene showed the fewest false alarms overall, which is consistent with its low actual confabulation rate.

Collectively, these results provide an overview of the gap between perceived and actual confabulation counts, and how they differ across different confabulation types. They also suggest factors which affect the perception-reality gap. First, the number of perceived confabulations tends to saturate under higher confabulation load (as in the city scene), which widens the perception-reality gap.
Second, immediate, verifiable spatial violations (SPE and RFA) are detected at higher rates, with visual feedback appearing as the primary error-detection channel. Finally, the availability of an reference state also affects the perception-reality gap. Target styles in the loft scene had no clear reference state, which results in a lower detection rate in interpretation confabulations compared to spatial confabulations which were easier to verify.



\section{Discussion}

Non-obvious findings from the study reveal several dynamics that inform design implications for confabulation mitigation beyond this specific workflow (RQ4). \textbf{First,} increasing confabulation frequency does \textit{not} proportionally increase user awareness. \Cref{sec:gap} demonstrated that perceived confabulation counts become saturated and uncorrelated with actual counts under high confabulation load, which demonstrates that \textbf{user vigilance is structurally unreliable as a detection mechanism} and mitigation must be \textit{system-side} rather than \textit{user-side}. \textbf{Second,} spatial failures are caught through natural perceptual awareness while semantic failures go more unnoticed (\Cref{sec:gap}). This suggests that the immersive medium creates an \textit{asymmetric detection landscape} that can be accommodated, but also \textit{exploited} by proposing \textbf{modality-specific confabulation mitigation strategies}. Third, task framing can also affect confabulation tolerance. For example, open-ended stylistic tasks in the loft scene reduced the number of perceived confabulations compared to spatial errors in constrained tasks like the office scene, which suggests that \textbf{the task framing itself can act as a mitigation lever}. Finally and most counter-intuitively, the fluency in LLM textual responses do not reflect response quality. \textbf{Confident and coherent output consistently masked incorrect actions}, as reported in the FSC, CMD, MID, and AEF confabulation types in \Cref{sec:taxonomy}. These findings jointly inform the following design implications.

\subsection{Design Implications}

\paragraph{DI1: Scene state should be re-verified against the LLM's prior claims before each interaction round.} Perceived confabulations saturate under high load (\Cref{sec:gap}) and confident LLM text cannot serve as the accurate record of the scene state (FSC-DONE, $n=28$). In such cases, neither the user nor the LLM can be relied upon to reflect the actual scene state and the correctness of the edit. Therefore, the system should re-ground each interaction turn with its actual scene state rather than relying on the model's belief or the user's belief.

\paragraph{DI2: Technical tokens should be flagged as out-of-vocabulary by default.} CMD-TOKEN ($n=20$) occurred because internal naming convention such as the ``\_st'' suffix is silently treated as a stylistic name by the system instead of a special identifier. Systems should clearly distinguish user-facing vocabulary from internal identifiers and flag uncertainty when a command references the latter.

\paragraph{DI3: Bind spatial commands to the real-time viewpoint instead of the global reference frame.} RFA-DIR ($n=30$) occurred from resolving egocentric terms (such as front/back) against a fixed global frame. Our study reinforces findings in reference frame ambiguity in prior literature~\cite{frank1998formal, chen2025envisionvr} and extends these findings by demonstrating that the `JSON-in, JSON-out' LLM-assisted workflow alone is ineffective when both the camera pose and global scene coordinates are passed to the LLM agent. The LLM often fails to provide responses based on the user's egocentric reference frame and additional mechanisms should be in place to provide this feature.

\paragraph{DI4: Provide a verifiable scope before bulk edits.} ESE-OVER/UNDER ($n=12$) clustered around commands with collective nouns. The system should not rely on the LLM to make a correct decision on the edit scope and should instead visually highlight objects within the edit scope to allow users to make corrections as appropriate.

\subsection{Limitations and Future Work}

This study was conducted in a controlled lab setting with a single system involved, which may limit the generalizability of findings. The sample of 24 participants limits the statistical power for type-level analyses. Future work will further evaluate whether the design implications reduce both the actual confabulation rates and the perception-reality gap in ecologically valid settings and extend the taxonomy to other LLM for XR workflows such as collaborative authoring and procedural content generation to establish a broader foundation for confabulation-aware design in XR.

\section{Conclusion}

This paper presents an empirical analysis of confabulations in LLM-assisted immersive scene editing. We contribute a taxonomy of nine confabulation types across three task contexts based on empirical interaction data, and provide an analysis of their prevalence, disruptiveness, and investigate the gap between their actual and perceived occurrence. We also provide a task-level analysis to discuss the different types of confabulations introduced by different task types, and how different task types result in different perceived user trust and overall load. The findings suggest that confabulations are not uniformly distributed across task types but cluster into spatial failures in design tasks with spatial constraints and semantically-driven failures in stylistic design tasks. Most importantly, the study reveals several counter-intuitive dynamics with broad implications for AI-assisted XR applications: user awareness of confabulations saturates under load and cannot be effectively relied on as a detection mechanism; the immersive nature of XR creates an asymmetry in perceived information across different modalities, which could be deliberately exploited for confabulation mitigation; task framing mediates confabulation tolerance which could also be leveraged as a mitigation lever; and confident LLM responses often mask incorrect actions, which motivates the need for independent validation checks for verbal output. Together, these findings show that confabulation mitigation in XR is not simply a matter of improving LLM accuracy, but rather a system-level design improvement. As LLM-assisted interaction becomes an increasingly foundational interaction paradigm for immersive environments, we hope this work provides a conceptual foundation and concrete design guidance for building robust and trustworthy systems.




\bibliographystyle{ACM-Reference-Format}
\bibliography{template}

@inproceedings{chen2024gaussianeditor,
  title={Gaussianeditor: Swift and controllable 3d editing with gaussian splatting},
  author={Chen, Yiwen and Chen, Zilong and Zhang, Chi and Wang, Feng and Yang, Xiaofeng and Wang, Yikai and Cai, Zhongang and Yang, Lei and Liu, Huaping and Lin, Guosheng},
  booktitle={Proceedings of the IEEE/CVF conference on computer vision and pattern recognition},
  pages={21476--21485},
  year={2024}
}

@article{chen2025envisionvr,
  title={EnVisionVR: A scene interpretation tool for visual accessibility in virtual reality},
  author={Chen, Junlong and Esparza, Rosella P Galindo and Garaj, Vanja and Kristensson, Per Ola and Dudley, John},
  journal={IEEE Transactions on Visualization and Computer Graphics},
  year={2025},
  publisher={IEEE}
}

@incollection{frank1998formal,
  title={Formal models for cognition—taxonomy of spatial location description and frames of reference},
  author={Frank, Andrew U},
  booktitle={Spatial cognition: An interdisciplinary approach to representing and processing spatial knowledge},
  pages={293--312},
  year={1998},
  publisher={Springer}
}

@inproceedings{sun2025layoutvlm,
  title={{LayoutVLM: Differentiable Optimization of 3D Layout via Vision-Language Models}},
  author={Sun, Fan-Yun and Liu, Weiyu and Gu, Siyi and Lim, Dylan and Bhat, Goutam and Tombari, Federico and Li, Manling and Haber, Nick and Wu, Jiajun},
  booktitle={Proceedings of the Computer Vision and Pattern Recognition Conference},
  pages={29469--29478},
  year={2025}
}

@article{shu2025gaussedit,
  title={Gaussedit: Adaptive 3d scene editing with text and image prompts},
  author={Shu, Zhenyu and Yu, Junlong and Chao, Kai and Xin, Shiqing and Liu, Ligang},
  journal={IEEE Transactions on Visualization and Computer Graphics},
  year={2025},
  publisher={IEEE}
}

@inproceedings{huang2024blenderalchemy,
  title={{BlenderAlchemy: Editing 3D Graphics with Vision-Language Models}},
  author={Huang, Ian and Yang, Guandao and Guibas, Leonidas},
  booktitle={European Conference on Computer Vision},
  pages={297--314},
  year={2024},
  organization={Springer}
}

@inproceedings{zheng2025editroom,
  title={{EditRoom: LLM-parameterized Graph Diffusion for Composable 3D Room Layout Editing}},
  author={Zheng, Kaizhi and Chen, Xiaotong and He, Xuehai and Gu, Jing and Li, Linjie and Yang, Zhengyuan and Lin, Kevin and Wang, Jianfeng and Wang, Lijuan and Wang, Xin},
  booktitle={International Conference on Learning Representations},
  volume={2025},
  pages={86791--86808},
  year={2025}
}

@inproceedings{de2024llmr,
  title={Llmr: Real-time prompting of interactive worlds using large language models},
  author={De La Torre, Fernanda and Fang, Cathy Mengying and Huang, Han and Banburski-Fahey, Andrzej and Amores Fernandez, Judith and Lanier, Jaron},
  booktitle={Proceedings of the 2024 CHI Conference on Human Factors in Computing Systems},
  pages={1--22},
  year={2024}
}

@article{chen2025llmer,
  title={Llmer: Crafting interactive extended reality worlds with json data generated by large language models},
  author={Chen, Jiangong and Wu, Xiaoyi and Lan, Tian and Li, Bin},
  journal={IEEE Transactions on Visualization and Computer Graphics},
  year={2025},
  publisher={IEEE}
}

@inproceedings{chen2025analyzing,
  title={Analyzing multimodal interaction strategies for llm-assisted manipulation of 3d scenes},
  author={Chen, Junlong and Grubert, Jens and Kristensson, Per Ola},
  booktitle={2025 IEEE Conference Virtual Reality and 3D User Interfaces (VR)},
  pages={206--216},
  year={2025},
  organization={IEEE}
}

@inproceedings{zhang2024vrcopilot,
  title={Vrcopilot: Authoring 3d layouts with generative ai models in vr},
  author={Zhang, Lei and Pan, Jin and Gettig, Jacob and Oney, Steve and Guo, Anhong},
  booktitle={Proceedings of the 37th Annual ACM Symposium on User Interface Software and Technology},
  pages={1--13},
  year={2024}
}

@inproceedings{tang2025llm,
  title={{LLM Integration in Extended Reality: A Comprehensive Review of Current Trends, Challenges, and Future Perspectives}},
  author={Tang, Yiliu and Situ, Jason and Cui, Andrea Yaoyun and Wu, Mengke and Huang, Yun},
  booktitle={Proceedings of the 2025 CHI Conference on Human Factors in Computing Systems},
  pages={1--24},
  year={2025}
}

@inproceedings{wu2023multimodal,
  title={Multimodal large language models: A survey},
  author={Wu, Jiayang and Gan, Wensheng and Chen, Zefeng and Wan, Shicheng and Yu, Philip S},
  booktitle={2023 IEEE International Conference on Big Data (BigData)},
  pages={2247--2256},
  year={2023},
  organization={IEEE}
}

@article{huang2023language,
  title={Language is not all you need: Aligning perception with language models},
  author={Huang, Shaohan and Dong, Li and Wang, Wenhui and Hao, Yaru and Singhal, Saksham and Ma, Shuming and Lv, Tengchao and Cui, Lei and Mohammed, Owais Khan and Patra, Barun and others},
  journal={Advances in Neural Information Processing Systems},
  volume={36},
  pages={72096--72109},
  year={2023}
}

@inproceedings{yang2025llm,
  title={Llm meets scene graph: Can large language models understand and generate scene graphs? a benchmark and empirical study},
  author={Yang, Dongil and Kim, Minjin and Mac Kim, Sunghwan and Kwak, Beong-woo and Park, Minjun and Hong, Jinseok and Woo, Woontack and Yeo, Jinyoung},
  booktitle={Proceedings of the 63rd Annual Meeting of the Association for Computational Linguistics (Volume 1: Long Papers)},
  pages={21335--21360},
  year={2025}
}

@article{huang2025survey,
  title={A survey on hallucination in large language models: Principles, taxonomy, challenges, and open questions},
  author={Huang, Lei and Yu, Weijiang and Ma, Weitao and Zhong, Weihong and Feng, Zhangyin and Wang, Haotian and Chen, Qianglong and Peng, Weihua and Feng, Xiaocheng and Qin, Bing and others},
  journal={ACM Transactions on Information Systems},
  volume={43},
  number={2},
  pages={1--55},
  year={2025},
  publisher={ACM New York, NY}
}

@article{smith2023hallucination,
  title={Hallucination or confabulation? Neuroanatomy as metaphor in large language models},
  author={Smith, Andrew L and Greaves, Felix and Panch, Trishan},
  journal={PLOS Digital Health},
  volume={2},
  number={11},
  pages={e0000388},
  year={2023},
  publisher={Public Library of Science San Francisco, CA USA}
}

@article{ji2023survey,
  title={Survey of hallucination in natural language generation},
  author={Ji, Ziwei and Lee, Nayeon and Frieske, Rita and Yu, Tiezheng and Su, Dan and Xu, Yan and Ishii, Etsuko and Bang, Ye Jin and Madotto, Andrea and Fung, Pascale},
  journal={ACM computing surveys},
  volume={55},
  number={12},
  pages={1--38},
  year={2023},
  publisher={ACM New York, NY}
}

@article{zhang2025llm,
  title={Llm hallucinations in practical code generation: Phenomena, mechanism, and mitigation},
  author={Zhang, Ziyao and Wang, Chong and Wang, Yanlin and Shi, Ensheng and Ma, Yuchi and Zhong, Wanjun and Chen, Jiachi and Mao, Mingzhi and Zheng, Zibin},
  journal={Proceedings of the ACM on Software Engineering},
  volume={2},
  number={ISSTA},
  pages={481--503},
  year={2025},
  publisher={ACM New York, NY, USA}
}

@article{zhang2025siren,
  title={Siren’s Song in the AI Ocean: A Survey on Hallucination in Large Language Models},
  author={Zhang, Yue and Li, Yafu and Cui, Leyang and Cai, Deng and Liu, Lemao and Fu, Tingchen and Huang, Xinting and Zhao, Enbo and Zhang, Yu and Chen, Yulong and others},
  journal={Computational Linguistics},
  volume={51},
  number={4},
  pages={1373--1418},
  year={2025},
  publisher={MIT Press 255 Main Street, 9th Floor, Cambridge, Massachusetts 02142, USA~…}
}

@article{sahoo2024comprehensive,
  title={A comprehensive survey of hallucination in large language, image, video and audio foundation models},
  author={Sahoo, Pranab and Meharia, Prabhash and Ghosh, Akash and Saha, Sriparna and Jain, Vinija and Chadha, Aman},
  journal={Findings of the Association for Computational Linguistics: EMNLP 2024},
  pages={11709--11724},
  year={2024}
}

@article{liu2024survey,
  title={A survey on hallucination in large vision-language models},
  author={Liu, Hanchao and Xue, Wenyuan and Chen, Yifei and Chen, Dapeng and Zhao, Xiutian and Wang, Ke and Hou, Liping and Li, Rongjun and Peng, Wei},
  journal={arXiv preprint arXiv:2402.00253},
  year={2024}
}

@article{bai2024hallucination,
  title={Hallucination of multimodal large language models: A survey},
  author={Bai, Zechen and Wang, Pichao and Xiao, Tianjun and He, Tong and Han, Zongbo and Zhang, Zheng and Shou, Mike Zheng},
  journal={arXiv preprint arXiv:2404.18930},
  year={2024}
}

@article{tonmoy2024comprehensive,
  title={A comprehensive survey of hallucination mitigation techniques in large language models},
  author={Tonmoy, SMTI and Zaman, SM and Jain, Vinija and Rani, Anku and Rawte, Vipula and Chadha, Aman and Das, Amitava},
  journal={arXiv preprint arXiv:2401.01313},
  volume={6},
  year={2024}
}

@inproceedings{li2023evaluating,
  title={Evaluating object hallucination in large vision-language models},
  author={Li, Yifan and Du, Yifan and Zhou, Kun and Wang, Jinpeng and Zhao, Wayne Xin and Wen, Ji-Rong},
  booktitle={Proceedings of the 2023 conference on empirical methods in natural language processing},
  pages={292--305},
  year={2023}
}

@article{xu2024hallucination,
  title={Hallucination is inevitable: An innate limitation of large language models},
  author={Xu, Ziwei and Jain, Sanjay and Kankanhalli, Mohan},
  journal={arXiv preprint arXiv:2401.11817},
  year={2024}
}

@article{zhou2023analyzing,
  title={Analyzing and mitigating object hallucination in large vision-language models},
  author={Zhou, Yiyang and Cui, Chenhang and Yoon, Jaehong and Zhang, Linjun and Deng, Zhun and Finn, Chelsea and Bansal, Mohit and Yao, Huaxiu},
  journal={arXiv preprint arXiv:2310.00754},
  year={2023}
}

@article{liu2023visual,
  title={Visual instruction tuning},
  author={Liu, Haotian and Li, Chunyuan and Wu, Qingyang and Lee, Yong Jae},
  journal={Advances in neural information processing systems},
  volume={36},
  pages={34892--34916},
  year={2023}
}

@article{zhu2023minigpt,
  title={Minigpt-4: Enhancing vision-language understanding with advanced large language models},
  author={Zhu, Deyao and Chen, Jun and Shen, Xiaoqian and Li, Xiang and Elhoseiny, Mohamed},
  journal={arXiv preprint arXiv:2304.10592},
  year={2023}
}

@article{omar2025multi,
  title={Multi-model assurance analysis showing large language models are highly vulnerable to adversarial hallucination attacks during clinical decision support},
  author={Omar, Mahmud and Sorin, Vera and Collins, Jeremy D and Reich, David and Freeman, Robert and Gavin, Nicholas and Charney, Alexander and Stump, Lisa and Bragazzi, Nicola Luigi and Nadkarni, Girish N and others},
  journal={Communications Medicine},
  volume={5},
  number={1},
  pages={330},
  year={2025},
  publisher={Nature Publishing Group UK London}
}

@article{dahl2024large,
  title={Large legal fictions: Profiling legal hallucinations in large language models},
  author={Dahl, Matthew and Magesh, Varun and Suzgun, Mirac and Ho, Daniel E},
  journal={Journal of Legal Analysis},
  volume={16},
  number={1},
  pages={64--93},
  year={2024},
  publisher={Oxford University Press UK}
}

@article{twist2025library,
  title={Library Hallucinations in LLMs: Risk Analysis Grounded in Developer Queries},
  author={Twist, Lukas and Zhang, Jie M and Harman, Mark and Yannakoudakis, Helen},
  journal={arXiv preprint arXiv:2509.22202},
  year={2025}
}

@article{peng2025understanding,
  title={Understanding and evaluating hallucinations in 3d visual language models},
  author={Peng, Ruiying and Li, Kaiyuan and Zhang, Weichen and Gao, Chen and Chen, Xinlei and Li, Yong},
  journal={arXiv preprint arXiv:2502.15888},
  year={2025}
}

@inproceedings{yang20253d,
  title={3d-grand: A million-scale dataset for 3d-llms with better grounding and less hallucination},
  author={Yang, Jianing and Chen, Xuweiyi and Madaan, Nikhil and Iyengar, Madhavan and Qian, Shengyi and Fouhey, David F and Chai, Joyce},
  booktitle={Proceedings of the Computer Vision and Pattern Recognition Conference},
  pages={29501--29512},
  year={2025}
}

@article{wang2024hallo3d,
  title={Hallo3d: Multi-modal hallucination detection and mitigation for consistent 3d content generation},
  author={Wang, Hongbo and Cao, Jie and Liu, Jin and Zhou, Xiaoqiang and Huang, Huaibo and He, Ran},
  journal={Advances in Neural Information Processing Systems},
  volume={37},
  pages={118883--118906},
  year={2024}
}

@inproceedings{hirzle2023xr,
  title={{When XR and AI Meet - A Scoping Review on Extended Reality and Artificial Intelligence}},
  author={Hirzle, Teresa and M{\"u}ller, Florian and Draxler, Fiona and Schmitz, Martin and Knierim, Pascal and Hornb{\ae}k, Kasper},
  booktitle={Proceedings of the 2023 CHI conference on human factors in computing systems},
  pages={1--45},
  year={2023}
}

@inproceedings{giunchi2024dreamcodevr,
  title={{DreamCodeVR: Towards Democratizing Behavior Design in Virtual Reality with Speech-Driven Programming}},
  author={Giunchi, Daniele and Numan, Nels and Gatti, Elia and Steed, Anthony},
  booktitle={2024 IEEE Conference Virtual Reality and 3D User Interfaces (VR)},
  pages={579--589},
  year={2024},
  organization={IEEE}
}

@article{kurai2025magicitem,
  title={{MagicItem: Dynamic Behavior Design of Virtual Objects With Large Language Models in a Commercial Metaverse Platform}},
  author={Kurai, Ryutaro and Hiraki, Takefumi and Hiroi, Yuichi and Hirao, Yutaro and Perusqu{\'\i}a-Hern{\'a}ndez, Monica and Uchiyama, Hideaki and Kiyokawa, Kiyoshi},
  journal={IEEE Access},
  year={2025},
  publisher={IEEE}
}

@inproceedings{hu2024scenecraft,
  title={{Scenecraft: An LLM agent for synthesizing 3D scenes as blender code}},
  author={Hu, Ziniu and Iscen, Ahmet and Jain, Aashi and Kipf, Thomas and Yue, Yisong and Ross, David A and Schmid, Cordelia and Fathi, Alireza},
  booktitle={Forty-first International Conference on Machine Learning},
  year={2024}
}

@inproceedings{yang2025llplace,
  title={LLplace: Embodied 3D Indoor Layout Synthesis Framework with Large Language Model},
  author={Yang, Yixuan and Lu, Junru and Zhao, Zixiang and Luo, Zhen and Dong, Wanxi and Sanchez, Victor and Zheng, Feng},
  booktitle={2025 IEEE/RSJ International Conference on Intelligent Robots and Systems (IROS)},
  pages={20685--20691},
  year={2025},
  organization={IEEE}
}

@article{zhao2026scenerevis,
  title={SceneReVis: A Self-Reflective Vision-Grounded Framework for 3D Indoor Scene Synthesis via Multi-turn RL},
  author={Zhao, Yang and Sun, Shizhao and Zhang, Meisheng and Shi, Yingdong and Yang, Xubo and Bian, Jiang},
  journal={arXiv preprint arXiv:2602.09432},
  year={2026}
}

@article{feng2023layoutgpt,
  title={{LayoutGPT: Compositional visual planning and generation with large language models}},
  author={Feng, Weixi and Zhu, Wanrong and Fu, Tsu-jui and Jampani, Varun and Akula, Arjun and He, Xuehai and Basu, Sugato and Wang, Xin Eric and Wang, William Yang},
  journal={Advances in Neural Information Processing Systems},
  volume={36},
  pages={18225--18250},
  year={2023}
}

@inproceedings{amershi2019guidelines,
  title={Guidelines for human-AI interaction},
  author={Amershi, Saleema and Weld, Dan and Vorvoreanu, Mihaela and Fourney, Adam and Nushi, Besmira and Collisson, Penny and Suh, Jina and Iqbal, Shamsi and Bennett, Paul N and Inkpen, Kori and others},
  booktitle={Proceedings of the 2019 chi conference on human factors in computing systems},
  pages={1--13},
  year={2019}
}

@article{romeo2026exploring,
  title={Exploring automation bias in human--AI collaboration: a review and implications for explainable AI},
  author={Romeo, Giuseppe and Conti, Daniela},
  journal={AI \& SOCIETY},
  volume={41},
  number={1},
  pages={259--278},
  year={2026},
  publisher={Springer}
}

@inproceedings{gurita2025breaking,
  title={Breaking Bad (Design): Challenging AI User Interface Accessibility Guardrails},
  author={Gurita, Alexandra-Elena and Vatavu, Radu-Daniel},
  booktitle={Proceedings of the Extended Abstracts of the CHI Conference on Human Factors in Computing Systems},
  pages={1--7},
  year={2025}
}

@article{bach2024systematic,
  title={A systematic literature review of user trust in AI-enabled systems: An HCI perspective},
  author={Bach, Tita Alissa and Khan, Amna and Hallock, Harry and Beltr{\~a}o, Gabriela and Sousa, Sonia},
  journal={International Journal of Human--Computer Interaction},
  volume={40},
  number={5},
  pages={1251--1266},
  year={2024},
  publisher={Taylor \& Francis}
}

@article{buccinca2021trust,
  title={To trust or to think: cognitive forcing functions can reduce overreliance on AI in AI-assisted decision-making},
  author={Bu{\c{c}}inca, Zana and Malaya, Maja Barbara and Gajos, Krzysztof Z},
  journal={Proceedings of the ACM on Human-computer Interaction},
  volume={5},
  number={CSCW1},
  pages={1--21},
  year={2021},
  publisher={ACM New York, NY, USA}
}

@article{zhang2025exploring,
  title={Exploring collaboration patterns and strategies in human-ai co-creation through the lens of agency: A scoping review of the top-tier hci literature},
  author={Zhang, Shuning and Wang, Hui and Yi, Xin},
  journal={Proceedings of the ACM on Human-Computer Interaction},
  volume={9},
  number={7},
  pages={1--43},
  year={2025},
  publisher={ACM New York, NY, USA}
}

@inproceedings{moruzzi2024user,
  title={A user-centered framework for human-ai co-creativity},
  author={Moruzzi, Caterina and Margarido, Solange},
  booktitle={Extended Abstracts of the CHI Conference on Human Factors in Computing Systems},
  pages={1--9},
  year={2024}
}

@article{zerilli2022transparency,
  title={How transparency modulates trust in artificial intelligence},
  author={Zerilli, John and Bhatt, Umang and Weller, Adrian},
  journal={Patterns},
  volume={3},
  number={4},
  year={2022},
  publisher={Elsevier}
}

@article{schmidt2020transparency,
  title={Transparency and trust in artificial intelligence systems},
  author={Schmidt, Philipp and Biessmann, Felix and Teubner, Timm},
  journal={Journal of Decision Systems},
  volume={29},
  number={4},
  pages={260--278},
  year={2020},
  publisher={Taylor \& Francis}
}

@inproceedings{shamsujjoha2025swiss,
  title={Swiss cheese model for ai safety: A taxonomy and reference architecture for multi-layered guardrails of foundation model based agents},
  author={Shamsujjoha, Md and Lu, Qinghua and Zhao, Dehai and Zhu, Liming},
  booktitle={2025 IEEE 22nd International Conference on Software Architecture (ICSA)},
  pages={37--48},
  year={2025},
  organization={IEEE}
}

@misc{Meta_2025, title={{Interaction SDK Overview | Meta Horizon OS Developers}}, howpublished = {\url{https://developers.meta.com/horizon/documentation/unity/unity-isdk-interaction-sdk-overview/}}, journal={Interaction SDK Overview | Meta Horizon OS Developers}, author={Meta}, year={2025}, month={August}, note = "Online, accessed 08-March-2026"}

@incollection{hart1988development,
title = {{Development of NASA-TLX (Task Load Index): Results of Empirical and Theoretical Research}},
editor = {Peter A. Hancock and Najmedin Meshkati},
series = {Advances in Psychology},
publisher = {North-Holland},
volume = {52},
pages = {139-183},
year = {1988},
booktitle = {Human Mental Workload},
issn = {0166-4115},
DOI = {https://doi.org/10.1016/S0166-4115(08)62386-9},
url = {https://www.sciencedirect.com/science/article/pii/S0166411508623869},
author = {Sandra G. Hart and Lowell E. Staveland}
}

@article{brooke1996sus,
  title={SUS-A quick and dirty usability scale},
  author={Brooke, John and others},
  journal={Usability evaluation in industry},
  volume={189},
  number={194},
  pages={4--7},
  year={1996},
  publisher={London, England}
}

@article{shneiderman1997direct,
  title={Direct manipulation vs. interface agents},
  author={Shneiderman, Ben and Maes, Pattie},
  journal={interactions},
  volume={4},
  number={6},
  pages={42--61},
  year={1997},
  publisher={ACM New York, NY, USA}
}

@inproceedings{feng2024large,
  title={Large language model-based human-agent collaboration for complex task solving},
  author={Feng, Xueyang and Chen, Zhi-Yuan and Qin, Yujia and Lin, Yankai and Chen, Xu and Liu, Zhiyuan and Wen, Ji-Rong},
  booktitle={Findings of the Association for Computational Linguistics: EMNLP 2024},
  pages={1336--1357},
  year={2024}
}

@article{saldana2021coding,
  title={The coding manual for qualitative researchers},
  author={Salda{\~n}a, Johnny},
  year={2021},
  publisher={SAGE publications Ltd}
}

\appendix

\end{document}